# PREDICTIVE PROTOTYPING: EVALUATING DESIGN CONCEPTS WITH GPT

Hilsann Yong
Singapore University of Technology & Design,
Singapore
hilsann_yong@mymail.sutd.edu.sg

Bradley A. Camburn
Singapore University of Technology & Design,
Singapore
bradley_camburn@sutd.edu.sg

**ABSTRACT**

*The design-build-test cycle is critical to realize innovative solutions. This cycle enables a team to evaluate a concept based on performance and iteration to enhance the design. However, testing is often time consuming and costly. While physics-based simulation, and strategic prototyping can substantially lower cost there are still significant investments required or potential limits in test scope until an integrated prototype can be produced. Recently, generative pretrained transformers (GPTs), have demonstrated strong performance in many high level reasoning and analysis tasks. Large language models (LLMs) or GPTs in particular, may have the potential to streamline the iteration cycle of generating meaningful designs insights more efficiently. This work explores whether a GPT can accurately predict information that would be gained during a prototyping effort such as cost, performance, and perceived usability. A novel approach is introduced to emulate design feedback using retrieval augmented generation (RAG) in conjunction with a GPT, specifically OpenAI's GPT-4o. This design method aims to bridge the gap between conceptual designs and physical. The method used in this paper leverages prototyping data scraped from the 'Instructables.com' database; thereby increasing the availability of relevant prototyping data to the model. Two efforts are reported. The first is a controlled study where predictions are made about a series of diverse designs. The GPT, and human designers were provided design sketches and asked to predict cost, performance, and usability. Performance of each condition is then compared to ground-truth physical prototyping results. A second effort reports on an experimental application, in which a physical prototype was produced based on recommendations from the GPT-RAG model. The performance of this prototype is compared against a baseline, commercial model, and a topology optimized model. The results indicate that the GPT-RAG predictions are more accurate than individual human or crowd estimations of cost and performance, while offering similar insights in terms of usability; the GPT-RAG inspired prototype also outperformed the commercial, and topology optimized prototypes. Interestingly this study also identified that repeatedly querying for cost and performance estimations from the GPT-RAG, and averaging the responses provided significantly more accurate results, highlighting that LLMs can emulate crowd behaviour, exhibiting the law of large numbers.*

## INTRODUCTION

Prototyping often predetermines a significant allocation of resource in the development and influences the probability of design success [1]. Therefore, to discover incompatible designs at a later stage of the design is far more detrimental in terms of time and cost as compared to during the initial design phase, shown in Figure 1 [2-4]. Extensive research indicates that 80% of a product's life-cycle cost is determined during the initial stages of design and development. Further reinforcing the need for cost-reduction strategies to be implemented earlier rather than as a post-process initiative [5]. This is a recurring theme across multiple design studies of strategic prototyping [6].

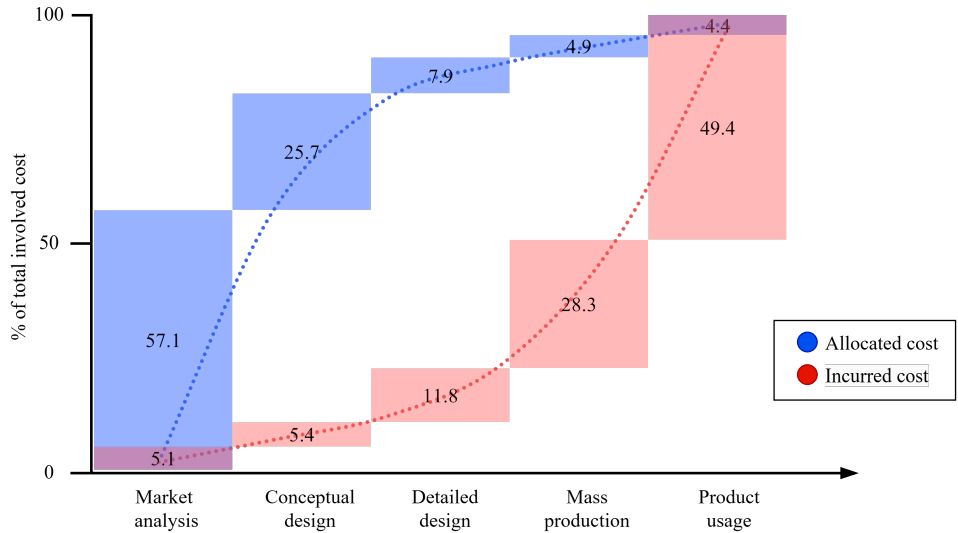

Figure 1: Plot of estimated gap between project allocated cost and actual incurred cost during the various design project phases

A design team may explore various prototyping strategies at different phases within a project, as depicted in Figure 2. They may be applied to prove design concepts, evaluate design alternatives and test for various features, for instance: functionality, usability, manufacturability, and desirability [7, 8].

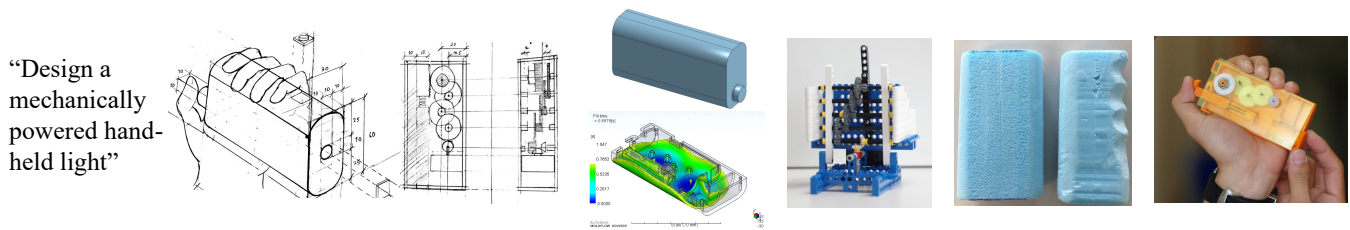

Figure 2: Palette of various prototyping efforts from a sample project

*Iteration*

Iteration is directly linked to performance increase in prototyping [9] and helps to yield more successful products, services, or systems [6, 10, 11]. It is a widely adopted concept in engineering that enables the refinement and eventual attainment of design requirements [12]. Typically, rather than pursue full-scale production designs, a team will produce a series of prototypes to evaluate design feasibility. Thought leading inventors like Thomas Edison and James Dyson failed substantially before achieving product success – Edison had 2774 attempts of numerous failed designs before arriving at the working design of the filament light bulb [13]. While Dyson had to develop 5127 prototypes before successfully delivering the DC-01 dual-cyclonic vacuum into the market [14]. These are two of the many pioneers that have popularised the idea of iteration. Moore's law is another famous testament to iterative enhancement [15]. Iteration is typically used to increase the probability of success for the market entry of consumer products – by means of ideation, creation and evaluation [8]. However, even though there is evidence that repeated iterations may result in diminishing costs [9] per test the effort and cost are linked to the complexity of the design in question and often substantial when higher fidelity testing is required [16]. Further, due to sunk costs and fixation effects, designers may hesitate to iterate once they have already produced an initial prototype, and, further, the more time spend on a single prototyping effort the more fixation effects occur [6, 17].

*Virtual Prototyping*

Research has shown that in the early phase of testing, physical and virtual prototypes offer similar benefits [18]. In principle, any digital information or interpretation created by computational methods are considered as virtual prototypes [7, 22]. A virtual prototype emulates some of the same characteristics of the design in question [7].

Designers employing virtual prototypes may pursue more iterations [19], furthermore, while virtual prototyping does not appear to result in a reduction of final prototype performance, but may still require substantial time [9]. This presents a notable gap between the iterative phase of design, build and test. Design engineers are often awaiting for an assessment from the simulation team before being able to make any critical design decisions [21].

*Low-fidelity Prototyping*

Another key dimension of prototyping, is between low-fidelity or high-fidelity. [18] Rapid, low fidelity prototyping offers some of the most valuable design insights [6]. Psychological research has shown that "low fidelity" prototypes benefit design teams in terms of transitioning failure into learning; provide a sense of progress and strengthen confidence in creative abilities [23]. Another paper also states that the idea of non-visual virtual prototypes aid designers to avoid design fixation, a challenge more prominent with physical prototypes []. Ultimately, prototyping as early as possible in the design process is often associated to improved overall design outcome [25], which is a key motivator for a new approaches to be continuously explored. Mockups, and low fidelity simulations, typically offer less accurate insights into factors like design performance or cost than higher fidelity alternatives.

*Motivation: The Need for a New Approach to Virtual Prototyping*

There is a spectrum between high and low fidelity virtual prototypes, just as seen in physical designs. Traditionally, engineers use a physics-based simulation approach to evaluate critical aspects of a design; and the team then transitions to subsystem and finally integrated system testing in later stages of development. Simulation requires effort to develop CAD models and specialized skillsets to develop a simplified model that will successfully run in a simulation environment. Designers must isolate key factors to make a viable simulation. For instance, designers must isolate internal volumes to model fluid flow, or isolate mechanical components to model stress propagation. Even after streamlining the model for simulation of a particular physical aspect, the corresponding simulations may be computationally intensive. While AI driven simulation approaches are reducing the associated computational complexity [26, 27], they still often require the same model preparation pipelines and only offer insights in a particular domain. Furthermore, it is also notoriously difficult to simulate human factors of a design such as desirability, or logistic factors such as predicted cost and performance.

In summary, physical prototyping requires a substantial amount of fabrication time and cost while virtual prototyping may often offer limited insights in specific physics domains. Thus, there is a need for a new virtual prototyping approach that offers rapid holistic insights about a specific design concept with minimal preparation time.

This research paper will address the potential of integrating GPTs, via retrieval augmented generation (RAG) with design repository data to accelerate a design engineer's design-build-test cycle (Figure 3). This GPT-based method is in effect a synthesis of several classical prototyping strategies: iterative, low-fidelity, and virtual prototyping. This approach is evaluated with a controlled experiment and a case study, to address the research questions:

1. Do GPTs provide accurate predictions of prototype performance and cost based on input design concept-sketches, as compared to ground truth physical prototyping?
2. Can GPTs accurately predict usability issues as compared to ground truth prototyping efforts?
3. How do GPT predictions compare to individual human, and crowd-based estimations of design cost, performance, and usability made from design sketches?
4. Does incorporation of a prototyping database significantly increase GPT prediction accuracy?

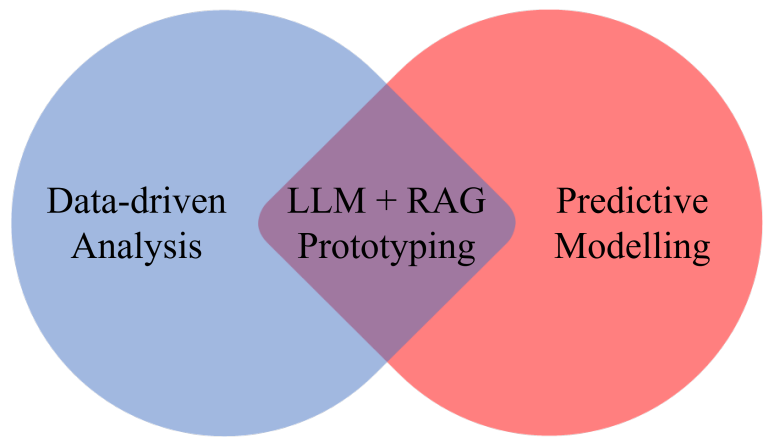

Figure 3: Predictive Prototyping with GPT-RAG

## METHODOLOGY

As Large Language Models (LLMs) become increasingly capable and accessible, they present an exciting opportunity to streamline design processes. The research explores how AI, particularly LLMs, can improve the design-build-test cycle in a design project. By focusing on mechanisms that directly support embodiment and evaluation, the method aims to leverage LLMs as a design prototyping environment. The eventual goal is to generate higher-quality design solutions through iterating based on insights from LLM evaluation. Hypothetically, by integrating meaningful and relevant datasets with the predictive capabilities of these models, users such as designers and engineers can generate valuable design insights more efficiently, consequently refining the iterative cycle and improving overall design quality.

*Design Method: Predictive Prototyping*

The proposed method involves a designer generating a conceptual design and evaluating its performance using a GPT model, which leverages on the relevant domain knowledge through Retrieval Augmented Generation (RAG). The primary objective is to accelerate the product development process by streamlining design evaluations through truncated iterations.

'Predictive Prototyping' as defined herein is the process of predicting the quality of a design based on a preliminary sketch or model in terms of likely results, if the design were prototyped, in terms of cost, performance, and usability. These factors are of paramount importance in deciding between or refining alternative design concepts. As there are a wide variety of sources used in training a GPT, we incorporate a dataset rich with information specifically relevant to prototype performances.

This is demonstrated using OpenAI's assistant platform, induced with a curated training dataset via RAG. By harnessing ChatGPT's Natural Language Processing (NLP) capabilities and extensive design knowledge from online repositories like 'Instructables.com', the GPT gains access to comprehensive and contextually relevant prototyping data to make predictions from. This enables the creation of outputs with a greater expected accuracy [28-30].

The approach is supported by findings that highlight a strong correlation between prototyping costs and the timing of implementation [31]. As shown in Figure 4, 'Predictive Prototyping' enables the user to obtain data analogous to that from a prototyping effort, but much earlier in the design process. Early-stage prototyping is more cost-effective, as late-stage iterations significantly increase incurred expenses. Virtual prototyping has proven particularly effective for validating design solutions [32], offering substantial reductions in both prototyping time and the costs of design modifications.

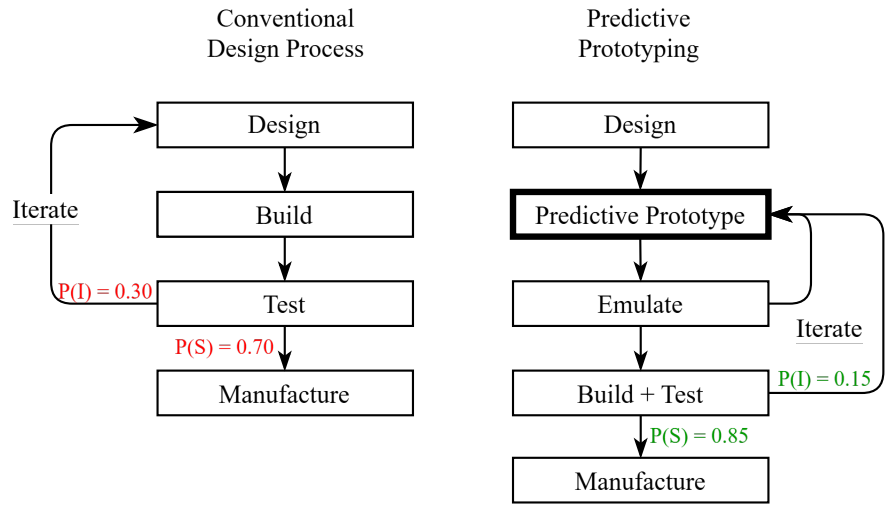

Figure 4: Comparison of the conventional design process vs. Predictive Prototyping; 'P(S)' = probability of success, 'P(I)' = probability of iteration

Predictive prototyping is intended to reduce the cost of iteration, and enable earlier acquisition of prototype feedback, which consequently improves the probability of success. In essence the design team can gain detailed information on likely cost, performance, and usability of a solution during the conceptual development stage, rather than much later in the process as is typical.

# EXPERIMENTAL EVALUATION

This is demonstrated This section outlines the experiments conducted to evaluate GPTs performance in predictive prototyping. The performance of GPT is compared to the outputs of 30 human participants with 30 generated contents by both ChatGPT ('GPT') and RAG-induced GPT ('GPT-RAG') across three fundamental aspects: cost, performance, and usability versus ground truth prototyping data. The subsequent section reports on a case study in which a given design is evaluated and refined based on feedback from the GPT. OpenAI's GPT-4o model, has been selected as the base model for comparison with the impact of incorporating the prototyping database.

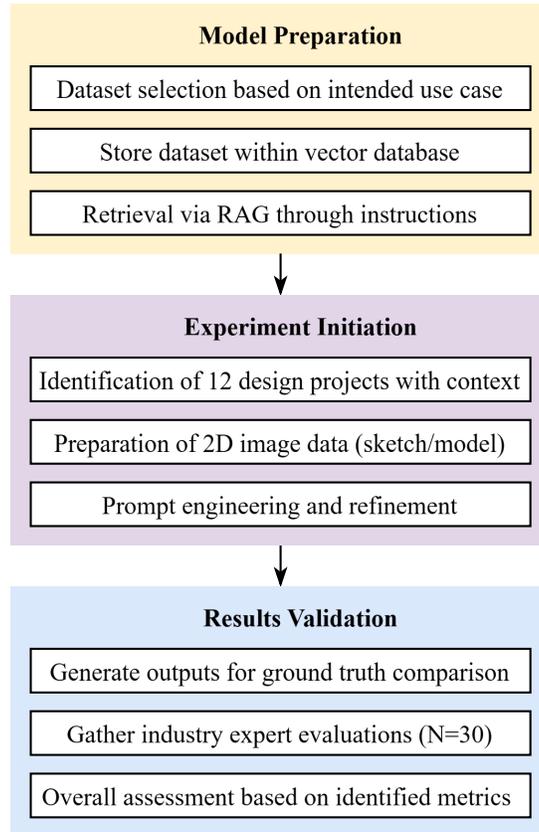

Figure 5: Overview of experimental methodology

*Model Preparation: Selected Database for RAG*
RAG is a technique that augments LLMs with external domain-specific knowledge sources [35, 36]. In an effort to mitigate hallucinations and improve model accuracy and relevancy [34], the dataset employed for RAG enhancement was sourced from 'Instructables.com', encompassing over 20,000 projects. For this study, the selected projects are extracted from the following categories: circuits, workshop, and craft [37].

Each project includes comprehensive documentation, featuring a bill of materials with cost breakdowns, performance expectations aligned with design requirements, constructive feedback for improvements and step-by-step instructional guides. This dataset provides structured, domain-specific knowledge, facilitating the transformation of abstract ideas into tangible outcomes. By integrating this database, it enables GPT-4o to offer more contextually rich feedback and design insights, supported by practical examples from actual user experiences [34, 38, 39].

*Experiment Initiation: 12 Design Evaluation Dataset*
A comparative study between industry expert responses and GPT outputs is conducted to quantify performance differences between humans and AI. For this research, 12 designs from various open-source design projects were selected based on the availability of detailed information, including cost-to-build, targeted performance, and design improvement feedback. These values are tabulated and established as ground truths or actual benchmarks that both humans and GPT models are expected to predict accurately. Designs listed in Table 1 have completed at least a full cycle of the design-build-test phase [34] and either have a sketch or model representation of the intended design.

The 12 designs are showcased in Table 1, where they have been curated from both online design repositories and research papers. Once the contextual data and 2D image are prepped for interpretation, advanced prompt engineering techniques have been incorporated to ensure that the generated outputs are relevant and accurate to the task.

For this research, the COSTAR framework was adopted due to its comprehensiveness – COSTAR: Context, Objective, Style, Tone, Audience and Response to ensure all bases are covered. In order to ensure that there is a fair comparison between all the various mediums, 'context' has decomposed further into the following three pointers: The problem, design solution, three key functions of the design and the physical dimensions, and weight if stated.

Central Limit Theorem indicates that repeated sampling results in closer estimates of a true value. As the responses of a GPT vary when repeatedly queried with the same question, we also explore whether repeated queries with averaging has any impact on accuracy. Crowdsourcing has been proposed in various design research efforts as a strategy for rapidly gaining accurate design insights. For the human prediction test, 30 industry experts were selected to participate in the experiment. For GPT and GPT-RAG testing, the models was queried 30 times with the same image and prompt; providing multiple outputs for each design evaluation.

While reviewing the variation in values, the following two factors will be considered:
- **Accuracy (Relevancy)**: The closeness of predictions relative to the ground truth.
- **Precision (Deviation)**: The span of variations in outputs, by standard deviation.

Table 1: Selected design projects for Human vs. AI evaluation with provided context, actual cost, and actual performance

| Design | Image of Virtual Prototype | Refined Context | Actual Cost (USD) | Actual Performance |
|---|---|---|---|---|
| 1 | | In emergency situations such as blackouts or night camping, portable and easy-to-use lighting is essential. This thumb-driven design features a mechanically powered emergency light that uses a rack-and-pinion system. When the rack is compressed, the gears transmit a driving force to rotate the DC motor shaft, thereby generating an electrical charge. During retraction, modified gears equipped with a clutch allow free rotation to return the system to its initial state. The generated charge is stored in a capacitor to power an LED, providing light when needed, such as during a power outage. | $19.80 | 50 lumens |
| 2 | | In mass casualty cases and resource-limited environments, access to a ventilator can be challenging. This solution provides an automated and portable method for delivering oxygen from a bag-valve mask. It is a mechanical ventilator that utilizes a pivoting cam arm to compress the bag-valve mask in a cyclic manner. Powered by a battery, it requires minimal power input, making it suitable for on-the-go use. This standalone operational machine is constructed using readily available components. | $310.00 | 750 ml/kg |

| 3 | 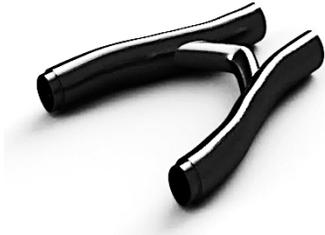 | This research examines the advantages of modern manufacturing techniques in achieving designs that are unattainable using conventional methods. A critical consideration is the minimum bending radius required for bending metal tubes. To address this challenge, additive manufacturing is employed to overcome bending limitations while preserving structural integrity and product quality. The study utilizes Selective Laser Melting (SLM) with Ti3-2.5 on a model optimized for Design for Additive Manufacturing (DfAM). The objective is to reduce part costs and optimize performance compared to a component manufactured using CNC methods. | $130.00 | 162 N/mm |
| --- | --- | --- | --- | --- |
| 4 | 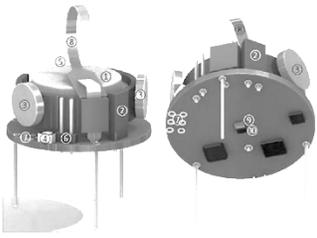 | Validating algorithms with collective robots has traditionally posed challenges for scalability. This product decentralizes cooperating robots, making scalability feasible through the strategic selection of components used in its construction. It can be assembled within five minutes and easily interacts with other units. This small robot features a simple design fitted with minimal viable components, such as light sensors and vibration-based locomotion, allowing it to navigate at a speed of 1 cm/s. The robot supports two key demonstrations: an orbit demonstration and a path-following demonstration. | $14.05 | ±2 mm |
| 5 | 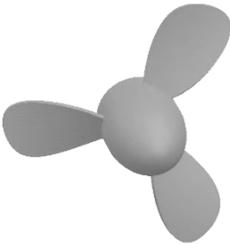 | The propeller is designed to generate thrust, enabling the vessel to move forward even against strong water currents. However, rapid prototyping does not accurately capture the actual performance of marine propellers. To address this, the design was optimized using an improved workflow that integrates simulation results with real-world conditions to predict the most effective design for fabrication. This design features a 3-blade propeller manufactured using the CNC method with brass, chosen for its corrosion-resistant properties. | $2,000.00 | 82.7 N |
| 6 | 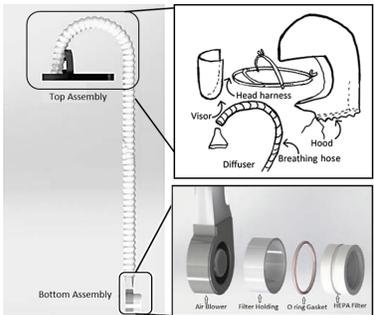 | The APF 25 wearable medical device features a top hood enclosure, supplying clean air through a tube connected to a blower equipped with a HEPA filter. The device is designed to operate within a temperature range of 20 to 45 degrees Celsius, helping to reduce water loss through perspiration. The noise level inside the hood is maintained at less than 80 dB, ensuring a comfortable experience. Additionally, the device can operate for up to six hours with minimal to no discomfort for the user. | $120.00 | 10 CFM |
| 7 | 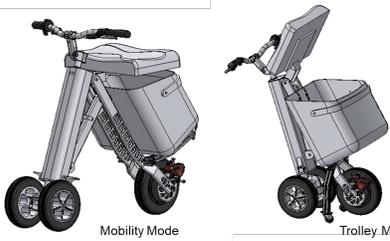 | Designed for daily tasks such as mobility and grocery shopping, this device aims to empower the aging population to function more independently in society. The mobility device features a dual function, enabling users to transport both themselves and cargo. It operates in two modes: mobility and trolley mode. The transformation process is straightforward, consisting of three easy steps that can be completed in just seven seconds. The device offers a cargo capacity of up to 25 kg for groceries and supports a user load of up to 100 kg. | $601.25 | 1.5h |
| 8 | 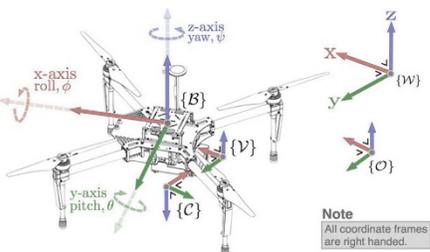 | Although micro aerial vehicles (MAVs) have seen significant advancements, challenges remain for precise tasks such as obstacle avoidance and precision agriculture. Research-grade MAVs offer superior solutions but are often expensive and challenging to maintain. This design presents a cost-effective, research-grade VTOL platform utilizing visual-inertial (VI) odometry, which combines lightweight construction, low power consumption, and advanced functionality for aerial robotics research. The platform incorporates an off-the-shelf visual-inertial sensor with a fisheye field of view of 133 degrees horizontally and 100 degrees vertically. The use of readily available components ensures a lightweight design, enabling reduced power consumption and extended flight times. | $4,489.25 | 3600g |
| 9 | 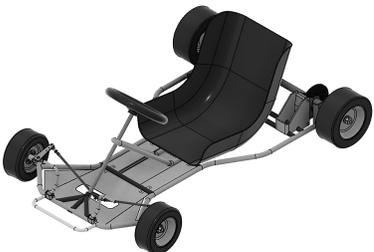 | This project aims to design an innovative, high-performance electric go-kart for both on-road and off-road use, pushing the boundaries of conventional design standards. Key features include an optimized size, chassis, suspension, and independently driven wheels. The front suspension incorporates unequal-length wishbones, with the top wishbone measuring 165 mm and the bottom wishbone 190 mm, along with mounting brackets offset by 20 mm to achieve a 9-degree caster angle. The go-kart is powered by two e-bike motors paired with compatible controllers and lithium batteries. Its construction features a thick plywood deck chassis and four 250 mm diameter wheels, offering a cost-effective and versatile vehicle. | $1,500.00 | 60 km/h |
| 10 | 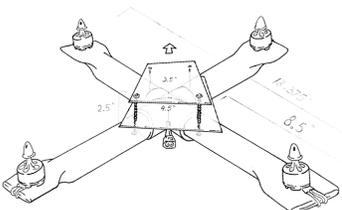 | The Ultimate PVC Quadcopter project focuses on creating a powerful, cost-effective quadcopter using durable PVC pipes for the frame.<br>The design is suitable for both beginners and experienced builders, offering an affordable and reliable solution for lifting cameras while achieving decent flight times. The use of 1" Schedule 21 PVC pipes and Lexan polycarbonate ensures durability and protection for the electronics. Through a series of sketches and prototypes, the quadcopter effectively meets these design criteria. | $300.00 | 11 mins |
| 11 | 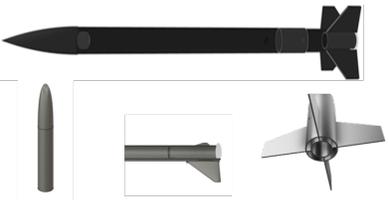 | This design study evaluates how 3D printing can enhance model rocket performance compared to traditional manufacturing techniques. By analysing existing rockets and iterating on 3D printed designs, we aim to match or surpass current models using data from the UK Rocketry Association under the D-class. The project involves testing rockets under consistent criteria, utilizing sensors for accurate data collection and validating results through simulations. The goal is to demonstrate how 3D printing can revolutionize model rocketry by reducing costs, simplifying construction, and improving performance in altitude, velocity, and acceleration. | $42.88 | 21.97 N |
| 12 | 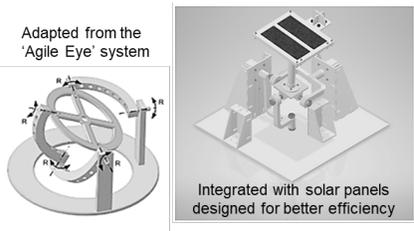 | Conventional solar panels often underperform due to their fixed positions, limiting their ability to capture light energy. Solar trackers can enhance efficiency by allowing panels to follow the sun's movement, enabling smaller panels to generate energy more effectively. This design incorporates a 2-degrees-of-freedom agile eye mechanism, a high-performance system capable of providing a camera with better range, velocity, and acceleration than the human eye. The compact design is tailored for personal consumers, making it suitable for use anywhere—both outside the home and within a garden or on a roof. | $120.00 | 40% |

*Result Validation: Three Dimensions of Evaluation*

In this pilot study, the research team ran a survey with 30 industry experts from two key domains: Design and Engineering. With participants spanning from 1 to 15 years of industry experience, there was an average 5.2 years of working experience across the participating industry experts. Among the 12 designs evaluated, each participant was tasked to estimate cost, predict performance, and identify at least 3 positive and 3 negative usability attributes of each design; each design was evaluated on multiple factors by 30 human participants.

*I. Cost Estimation & Performance Prediction*

As the cost estimation and performance prediction are straightforward to evaluate with the metrics of quantifiable measures, this enables the approach of through a four-prong data analytics approach of: Approximation error in percentage, Root Mean Square Error (RMSE), Standard Deviation (SD) and Analysis of Variance (ANOVA) for assessment of statistical significance.

The approximation error is intended to showcase a quick comparative view of the distance between the predicted data point and the ground truth value, in percentage units. These collected data points are plotted in figures 6 and 8, differentiated by colour and symbol shapes: grey triangles (GPT), blue diamonds (GPT-RAG) and orange squares (Industry experts). The centerline is indicated at 0%, which represents the 'ground truth' values extracted from the 12 design project descriptions. The y-axis are the 12 various designs evaluated, and the x-axis represents the deviation from the actual value (ground truth). Essentially, if a dataset mean is closer to the centerline, this indicates it is more accurate than a mean that is further away.

To further exhibit the key characteristic of the dataset, the error margin is calculated through the RMSE as a form of the loss function for the model. This is evaluated by individually calculating the difference between the actual and predicted error values for each data point. The resultant errors are then squared to remove any sign convention, summed up, and divided by the sample size (N=30). Finally, the square root of this value is the result of each point for the specified designs. These tabulated values are plotted in figures 7 and 9, to showcase the regression line, across the 12 designs. If the line is parallel and closer to the x-axis, it indicates that the model is more accurate and precise as compared to the designs further from the x-axis.

Finally, to ensure that the variation between each group of datasets is statistically significant, the one-way ANOVA test was conducted to determine if the p-values are lesser than 0.05; These values are tabulated into Table 3 between cost, performance and the influence of single querying and multiple querying.

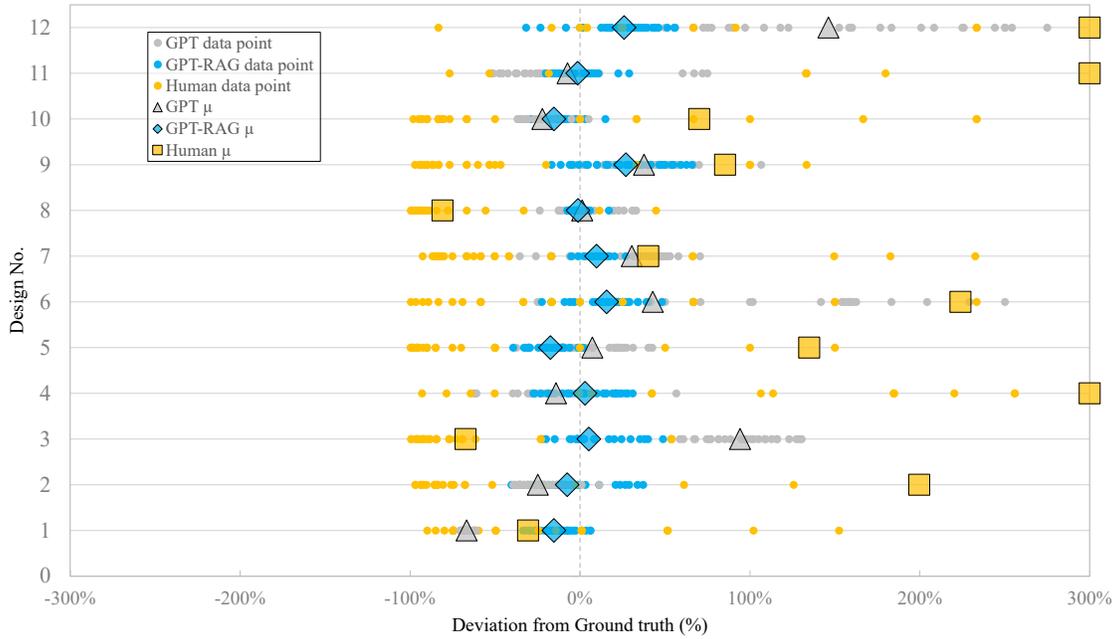

Figure 6: Approximation error of cost estimation in % from the mean across GPT, GPT-RAG and humans; Design details are listed above in Table 1

From Figure 6, it showcases that GPT-RAG, representative of 'Predictive Prototyping', has demonstrated superior accuracy in 11 out of the 12 evaluated designs, outperforming both GPT and human assessments in terms of proximity to the actual values. The GPT-RAG model estimations showcase a distinctively narrow spread across 30 iterations in comparison to its competitors, indicating high levels of both accuracy and precision.

The predictive capabilities of the model can be attributed to its integration with readily available cost data from the online design and prototyping repository, 'Instructables.com'. These comprehensive data repositories provided the model with a reliable reference framework, improving its ability to reproduce accurate cost predictions. In contrast, GPT and human participants lacked prior knowledge to access such extensive cost-specific information, resulting in comparatively less precise estimates.

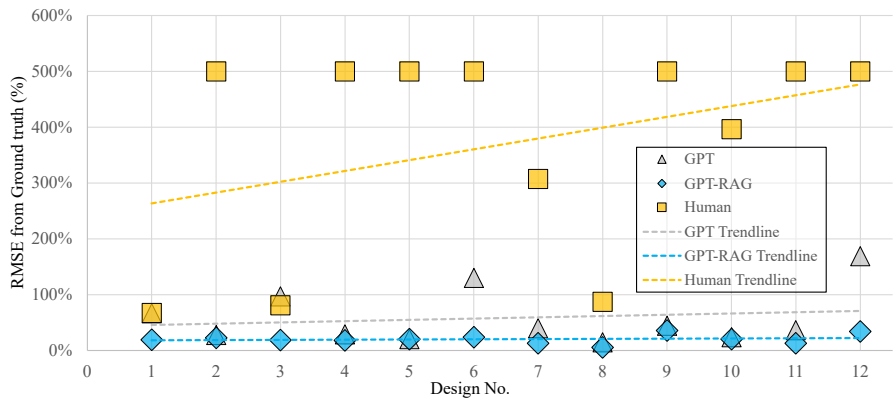

Figure 7: RMSE trendline comparison across GPT, GPT-RAG and human point estimates for cost estimation

These findings, which are further supported by the GPT-RAG trendline in Figure 7, demonstrates the efficacy of 'Predictive prototyping' as a tool for evaluating design costs. This approach of utilizing domain-specific data to rapidly provide accurate predictions, can help accelerate design engineering workflows by offering stakeholders actionable insights early in the process, minimizing potential redesigns, and reducing reliance on costly physical prototypes.

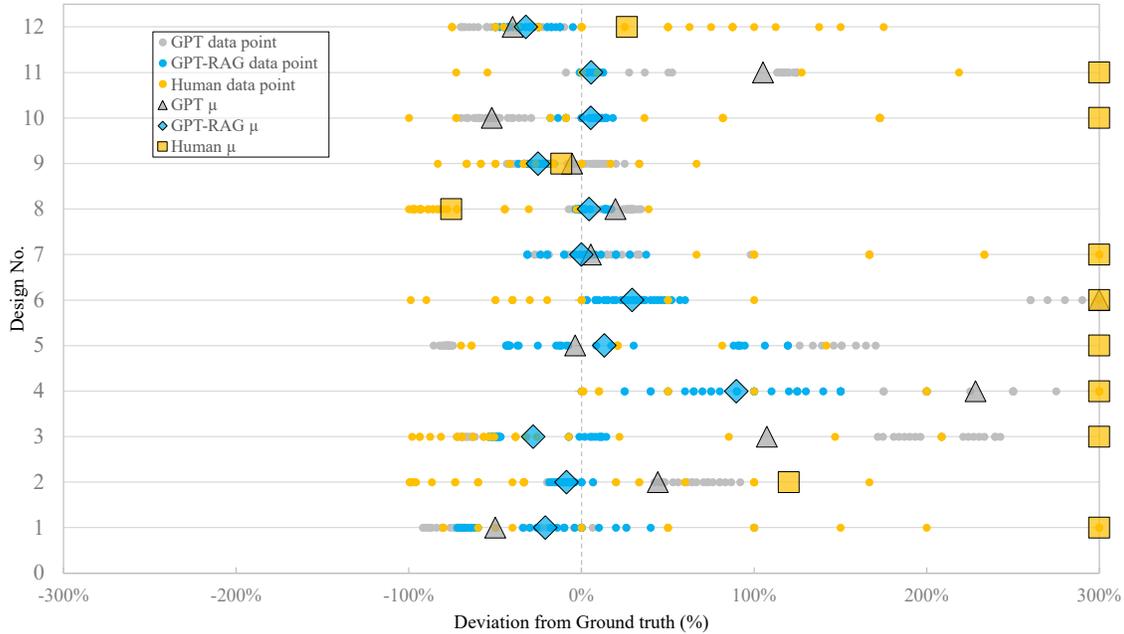

Figure 8: Approximation error of performance prediction in % from the mean across GPT, GPT-RAG and humans

In the study of performance predictions, Figure 8 compares predictions across the three same groups, but with different datasets. The prompt induced was specific to product performance and the measurable metric. In this experiment, GPT-RAG consistently demonstrated superior accuracy in 11 out of the 12 evaluated designs, outperforming both GPT and human assessments in terms of proximity to the actual values. However, in contrast to Figure 6, there is a noticeable wider spread of data points with significantly less precision in relation to the ground truth values. This increased variability is attributed to the contextual nature of product performance metrics, where a single design may serve multiple use cases.

For instance, a handheld light may have varying performance requirements: Emergency scenarios may prioritize lower lumens (30–50 lm) for localized visibility, while for recreational use, such as night camping, higher lumens (300–500 lm) are needed for navigation. This variability highlights the challenges of performance prediction and the importance of the user's contextual understanding when evaluating product functionality.

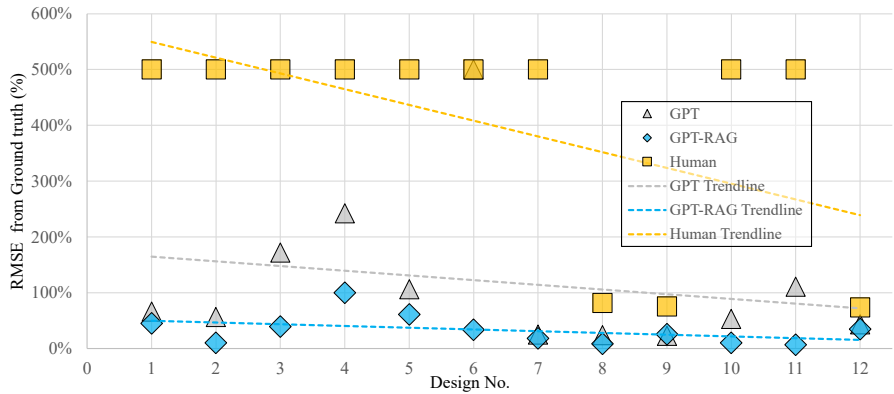

Figure 9: RMSE trendline comparison across GPT, GPT-RAG and human point estimates for performance prediction

On further investigation, as shown in Figure 9, the RSME trendline shows that human evaluators have a higher likelihood to over predict performance, which is likely driven by the desire of designers and engineers to overengineer solutions as a form of safety factor, which comes into play when considering manufacturing tolerances and real-world scenarios. This further demonstrate the efficacy of 'Predictive prototyping' as a tool for evaluating design performance, when induced with relevant data and insights.

Table 2: Accuracy and precision comparisons between single and repeated querying for GPT, GPT-RAG and humans

| Accuracy Metrics | GPT Single Query | GPT Multiple Query | GPT-RAG Single Query | GPT-RAG Multiple Query | Human Individual | Human Group |
|---|---|---|---|---|---|---|
| Cost – RMSE | 58% | 45% | 20% | 12% | 194% | 85% |
| Cost – Average Accuracy | 8/12 (67%) | 8/12 (67%) | 12/12 (100%) | 12/12 (100%) | 0/12 (0%) | 4/12 (33%) |
| Cost S.D. | 49% | 48% | 8% | 9% | 117% | 64% |
| Performance – RMSE | 109% | 86% | 33% | 22% | 385% | 211% |
| Performance – Average Accuracy | 4/12 (33%) | 7/12 (58%) | 10/12 (83%) | 11/12 (92%) | 0/12 (0%) | 4/12 (33%) |
| Performance S.D. | 109% | 111% | 27% | 24% | 365% | 192% |

Table 2 provides a summary of the experimental results consisting of RMSE, average accuracy and precision for both cost and performance across GPT, GPT-RAG and human evaluators. This provides an overview of the impact of RAG-induced GPT and multiple querying. A bandwidth of 50% has been agreed within the research team, to be considered as an accuracy tolerance which is incorporated into the average accuracy: This means that if the RMSE falls within 50% of the ground truth value, it will be calculated into the average accuracy (out of 12 designs).

As a reference, the lower the error indicated by the RMSE value, the better the performance of the model. While the higher the accuracy score, the better the prediction capabilities of the model. Lastly, the SD showcases the spread of the data, which showcases how precise are the predictions as a bandwidth across the dataset.

To better discern between single and multiple querying, the research team has established for individual or single query values, to utilize the RMSE of every individual data point across the 30 inputs in each group. In contrast, for multiple querying the mean absolute error (MAE) of each dataset is used as a representative for the single point estimate of the group.

Based on observation, GPT and GPT-RAG shows that multiple querying does generally improve average accuracy but has little influence on SD values. However, the improvement was more evident for human evaluators – The introduction of crowdsourcing results in 'casting of a larger net' to capture more variation in values. This is increases the odd of being closer to the actual value, essentially improving the accuracy of the group, in this case from 194% to 85% for cost RMSE and 385% to 211% for performance RMSE. This huge variation in performance prediction is likely due to the misinterpretation of human evaluators for the use case of design 11, as an example – a home-made model rocket optimized via 3D printing, some participants mistakenly assumed it was an actual space rocket, resulting in the inflated cost and performance estimates. Despite the physical dimensions being provided, this misinterpretation illustrates the limitations of human evaluators, who may fail to correctly contextualize the provided data.

There is a recurring trend across all three groups that multiple querying does help improve the accuracy, especially human evaluators where initially there was 0% for both cost and performance in terms of average accuracy, but with crowdsourcing the accuracy improved to 33%. Both GPT and GPT-RAG were outperforming human evaluators in this aspect, which is likely due to cost and performance being quantitative measures that is fundamentally based on data.

In conclusion, across the three different groups: GPT, GPT-RAG and Humans, GPT-RAG is consistently superior to its competitors, with a much better accuracy and narrower bandwidth of deviations. This provides confidence that the introduction of RAG for 'Predictive prototyping' enables the model to be more relevant for design evaluation. Table 3 summarizes the p-values calculated across the 4 scenarios with the one-way ANOVA test. All p-values are statistically significant.

Table 3: One-way ANOVA test for statistical significance across the three different groups

| GPT vs GPT-RAG vs Humans | Query Type | p-value | Statistically significant? ($p < 0.05$) |
|---|---|---|---|
| Cost | Single | <0.001 | *Yes* |
| Cost | Multiple | 0.002 | *Yes* |
| Performance | Single | 0.001 | *Yes* |
| Performance | Multiple | 0.003 | *Yes* |

This quantitative study has provided a foundation for extending the methodology to include the evaluation of design usability, a qualitative metric for achieving a comprehensive assessment of successful products. In the following experiment, given GPT-RAG's demonstrated superiority over GPT in the cost and performance predictions, only GPT-RAG will be considered in predicting usability issues.

*II. Design Usability*

In this study, design usability is measured by predicting non-obvious issues that usually arises at a later stage of the design process, issues that typically identified by subject matter experts during design reviews or during mass product usage. Although there are tools such as Failure Mode and Effects Analysis (FMEA) that support this process, it is often time-consuming and ineffective for designers working with tight timelines. Furthermore, there is a potential risk that designers may overlook potential concerns if the build and test planning is not meticulously executed. Therefore, the goal of this approach is to streamline the iterative design process by offering a more effective method for risk management, which helps in early identification of potential issues.

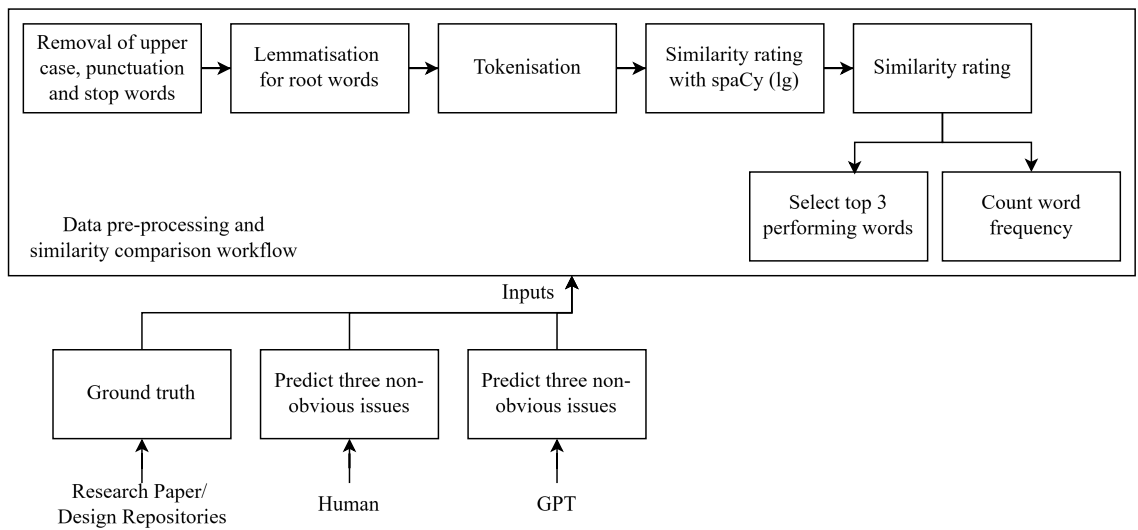

Figure 10: Workflow for decomposing ground truths and identifying issues by Human and GPTs

Taking inspiration from the System Usability Scale (SUS), this experiment adopts this universal approach for qualitative assessment, tasking both participants and GPT with predicting three positive aspects and then three potential issues for each given design. Structuring the queries in this order helps to facilitate effective brainstorming for points of failure that may not be immediately apparent.

The research team leverages on open-source NLP tools like spaCy's large english model, that has been trained on lexical databases such as 'WordNet'. shown in Figure 10 is the workflow that commences based on three inputs: The ground truth extracted from the project description, alongside the generated output from GPT and human evaluators. As the initial task was for participants to predict 3 to 5 non-obvious issues, these were generated in open-ended sentences. The sentences were further decomposed and lemmatized into key root words for pairwise comparison. Finally, across each of the 12 evaluated designs, the top 3 words with the highest similarity score is selected as a representative value. Some examples of keywords include "integrated", "modular" and "portable".

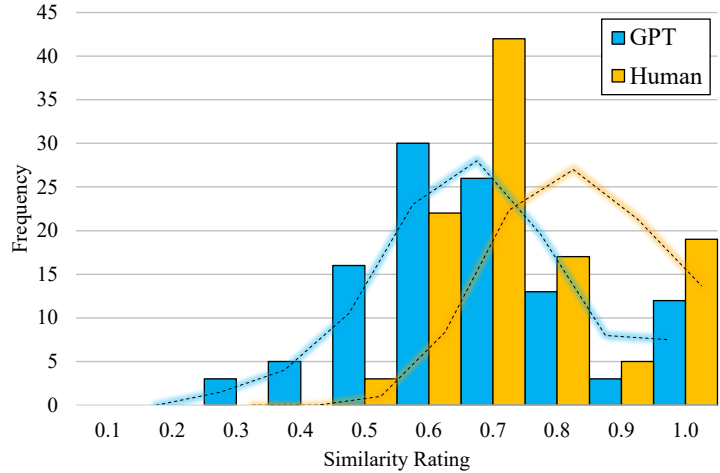

Figure 11: Human vs. AI prediction of non-obvious usability issues, versus ground truth prototyping data

With the top 3 performing words and their frequencies tabulated, the graph on Figure 11 is plotted with the y-axis showing token frequencies and the x-axis representing similarity ratings. An estimated 312 words were reviewed and distributed by similarity rating. The results indicate that human responses achieved a higher overall similarity score (81.2) compared to GPT (72.2). Based on t-testing between Human and GPT, $p < 0.001$ and therefore this analysis is considered as statistically significant.

While human responses generally achieved higher similarity scores to issues identified in the original design documentation sources, AI-generated predictions demonstrated a broader range of predictions, which reflects the capacity of the model for generalization across diverse scenarios. Yet, despite this variability, GPT still displays sufficient accuracy in identifying non-obvious issues, particularly those less frequently addressed by humans. This highlights its potential to complement human users by uncovering additional design considerations that may otherwise be overlooked.

*Conclusion & Discussion*

Drawing upon multiple studies comparing large language models (LLMs) and human evaluators [1, 40-44], this design methodology emphasizes on the value of leveraging AI as a tool for accelerating the design process. Through a data-driven approach, this research highlights the integration of three key dimensions for design evaluations – Cost, performance, and usability. In this comparison, GPT is representative of 'Predictive prototyping'.

Table 4: Overall comparison between GPT vs Human, across 12 designs based on multiple query and crowdsourcing

|  | GPT | Human |
|---|---|---|
| Cost | 67% | 33% |
| Performance | 58% | 33% |
| Usability | 72% | 81% |

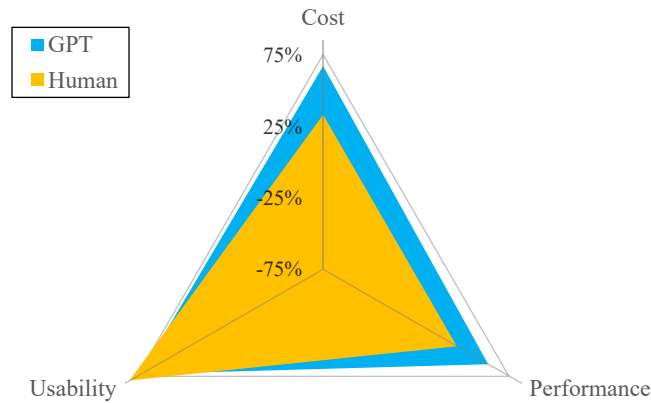

Figure 12: Overall performance of Human vs. AI across the three key metrics

Table 6 and Figure 12 highlights that while GPT excels in data-relevant tasks such as cost and performance evaluations, human evaluators outperform AI in usability assessments, such as identifying non-obvious issues rooted in real-world experiences. This complementarity presents an opportunity for human-AI collaboration, harnessing their respective strengths to achieve more accurate and efficient designs at an accelerated pace.

As design evaluation plays an essential role in the product development process [41], tools like 'Predictive prototyping' can help bridge knowledge gaps between designers and optimized designs. This form of virtual prototypes enables economical design evaluation with higher flexibility enriched by data. In retrospect, the growing number of emerging research work pertaining to the curation of domain-specific datasets for model fine-tuning, provides a glimpse into the exciting future of data accessibility [45-48].

Over the past four decades, research has demonstrated that product success is driven by multidimensional factors that contribute to long-term customer satisfaction and competitive advantage [49, 50]. By adopting this multidimensional framework, this research reinforces the critical role of factors like cost, performance, and usability in achieving sustainable product success. The proposed conceptual model, focusing on these independent variables, reflects the relevance of these dimensions in successful design practices, particularly within industries striving to remain competitive in an increasingly AI-driven market.

In conclusion, this paper emphasizes the potential of AI-powered design evaluation as an indispensable tool for the future of product development. As predictive prototyping and crowdsourcing continue to evolve, the synergy between human insight and AI capability will continuously enable better product design processes.

**CASE STUDY**

This case study explores a design challenge focused on enhancing the performance of a hairdryer attachment. Traditional attachments often restrict airflow, causing high-pitched whistling sounds during use which compromises user experience. The primary objective was to redesign the attachment to optimize airflow and mitigate acoustic issues, using the least material as possible. An indicator of success was defined as the new design surpassing the performance of the original, a critical aspect in new product development to secure a competitive market edge and deliver an optimized user experience.

To ensure the reliability of predicted performance improvements, the team conducted a physical prototype study to validate the results of the initial predictive analysis and whether this can support iterating on the design to enhance performance. While working with existing designs excluded from the sample, the study demonstrated the potential of this innovative and effective approach to tackle a complex real-world problem.

*Nozzle Design Permutations*

In the context of the hairdryer attachment, the nozzle was identified as the key component influencing overall performance. To optimize its functionality, various design permutations were explored and evaluated for their impact on airflow and acoustics performance. The research team utilized advanced design tools, including Autodesk Fusion's generative design features and GPT-RAG assisted techniques, to optimize the nozzle design.

Figure 13 illustrates the range of design permutations explored. Design A represents the baseline, developed to meet visual and production requirements and serving as the foundation for the existing design. Design B, created using GPT-RAG through an interactive chat interface, incorporated both contextual and pictorial inputs to retrieve relevant data from the vector store. This approach builds upon the baseline design, representing an evolution by Human-AI collaboration. Lastly, Design C, generated through Autodesk Fusion's generative design tool, was derived by the basic geometric components of the baseline design (e.g., rectangular outlet block, circular inlet hole) and predefined constraints for the algorithm to identify an optimized configuration based on requirements. In this scenario, a fluid path generative design was executed with the objective of minimizing pressure drop and reducing part weight. The pressure drop is determined based on the initial measured performance of the off-the-shelf supersonic.

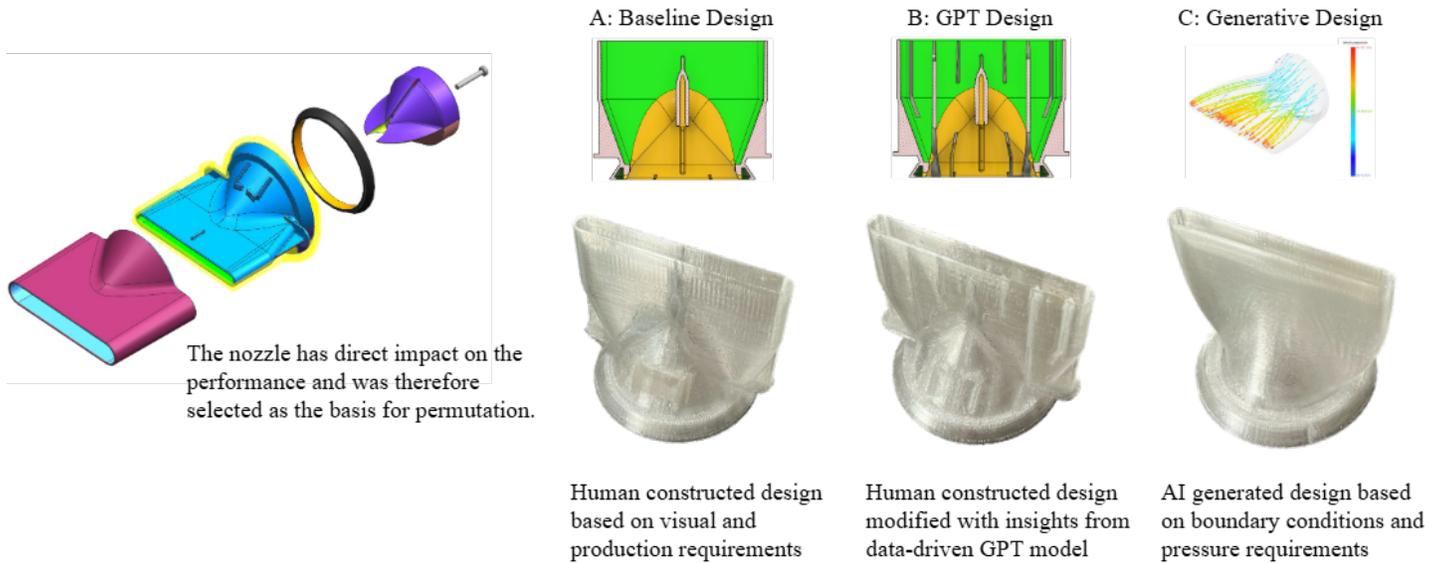

Figure 13: Comparison of various nozzle design permutations

*Comparative Test Setup*

The overall performance improvement was evaluated using two key metrics that directly influence user experience: airflow rate (m/s) and acoustic performance (dB). Fixed variables included the position and mode of the hair dryer, while independent variables encompassed different attachment design permutations and the anemometer's position, which ranged from 0.05 m to 0.25 m, as illustrated in Figure 14. This range replicates typical user scenarios and accounts for the potential impact on nearby individuals. Performance testing was conducted in a temperature-controlled, noise-isolated environment to ensure consistent conditions. This controlled setup allowed any observed performance improvements to be attributed solely to the design modifications, eliminating the influence of external factors.

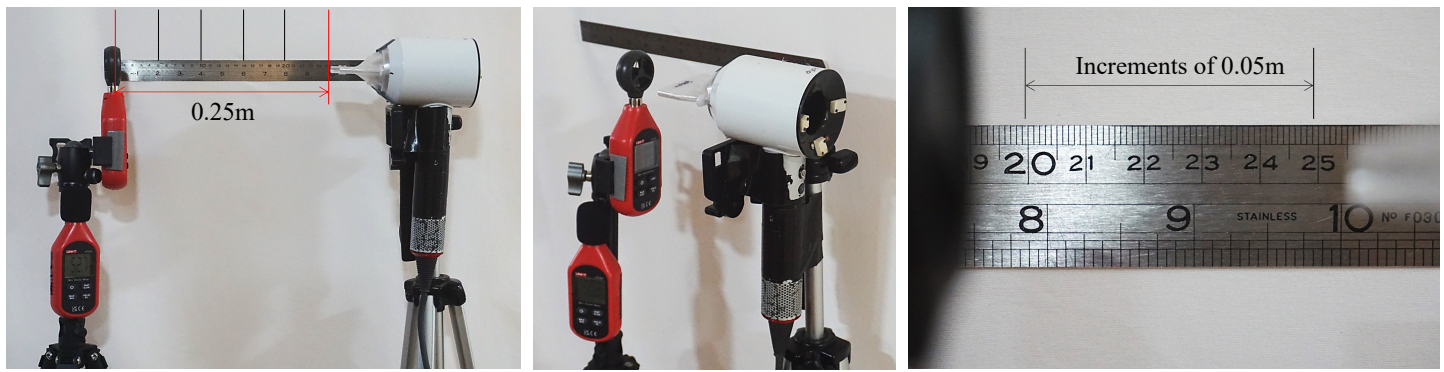

Figure 14: Test setup for airflow and acoustic performance measurements; Readings were taken across 0.25 m in 0.05 m increments using an anemometer and sound meter

*Results and Discussion*

Upon completion of the test, the results showcase a distinctive difference between each of the design permutations. As reflected in Figure 15, the y-axis represents the airflow in m/s, while the x-axis represents the increments in distance between the nozzle outlet and the anemometer in cm. Each data point is accompanied by the size of the bubble which represents the acoustic performance; this means that the larger the size of the bubble, the louder the induced sound in decibels. As a reference, the best average acoustic performance of the generative design is at 74.7dB while the worst performing is the baseline design is at 77.3dB – Considering that sound is on a logarithmic scale, this difference is exponential and can be significantly affect the user experience.

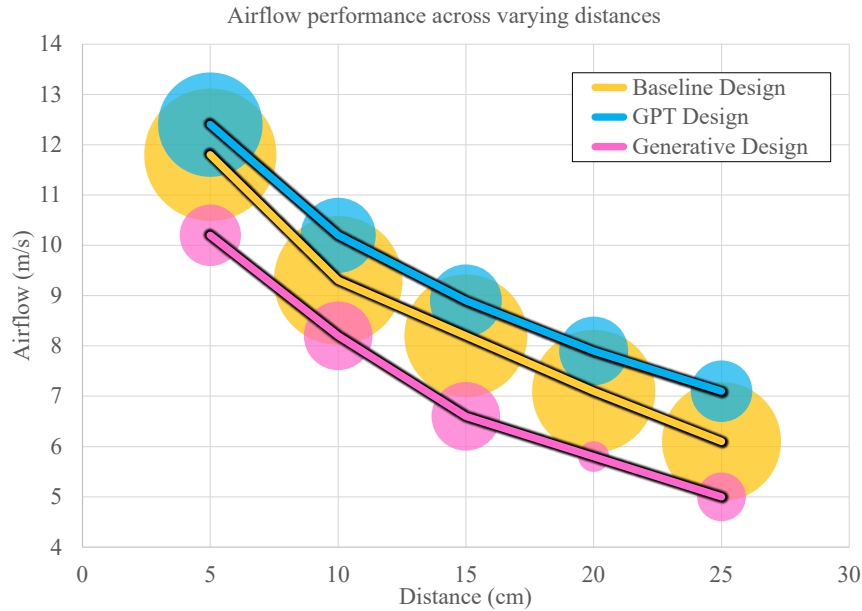

Figure 15: Airflow measurements, with bubble size indicating acoustic performance at each point (smaller is better for acoustics)

As shown above, the baseline design for benchmarking achieved a peak airflow of approximately 11.8 m/s at 0.05 m. This outperformed the generative design, which experienced a drop in performance with a peak airflow of 10.2 m/s, though it offered a significantly improved acoustic performance with a 3 dB reduction. The top-performing design, the GPT-RAG design, reached a peak airflow of 12.4 m/s, resulting in a 5% improvement in airflow and a 2% improvement in acoustic performance. However, based on the average across varying distances, there was an overall 10% improvement in airflow performance and a 3% improvement in acoustic performance. These results are summarized in Table 3, which compares the predicted values with the actual outcomes.

Table 5: GPT-RAG verbatim suggestions for improving airflow and acoustic performance; Research team incorporated these into the baseline design for testing, which showed a 10% improvement in airflow, as predicted, but fall short to a 3% improvement in acoustics

| | Input Prompt | Generated Output | Actual Improvement |
|---|---|---|---|
| Details | 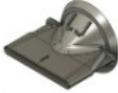 "Based on the given image, improve the performance of the provided hair dryer attachment design. How would you modify this hair dryer attachment design to perform better in terms of airflow?" | "Airflow Enhancers (Vanes/Fins): Incorporate internal vanes or fins that help direct the airflow more efficiently. These can guide the air in a laminar flow pattern, reducing turbulence and increasing airflow.<br><br>Thickness: 1.0 - 1.5mm<br>Height: 6.0 - 10.0mm<br>Reinforcement: 0.5 - 1.0mm thick ribs<br>Location: Near the inlet & outlet inside the airflow pathway<br>Main channel: 3-4 vanes<br>Outlet section: 6-8 vanes<br><br>An expected improvement of 10-15% can be expected." | 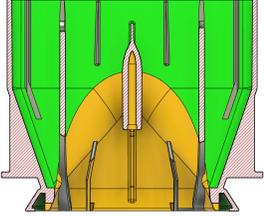 |
| Airflow (m/s) | - | 10 to 15% | **10%** |
| Acoustic (dB) | - | 6 to 12% | **3%** |

The results demonstrated in Figure 15, based on the suggestions identified in Table 3, highlight the potential benefits of using AI-driven tools, such as Predictive Prototyping, to generate superior design quality outputs. These findings are derived from an average of three steady state runs per data point, ensuring repeatability and reliability. While these results are promising, there are limitations to this method which will be addressed in the following section.

**CONCLUSION AND LIMITATIONS**

In summary, 'Predictive Prototypes' is a data-driven tool designed to bridge the gap between domain-specific knowledge and GPTs, facilitating useful applications. The paper outlines the relevant steps to leverage an open-source design repository (Instructables.com) and embed it into a vector datastore for Retrieval-Augmented Generation (RAG). The experimental results highlight the advantages of predictive prototyping, as evaluated using the metrics most relevant to both designers and engineers: cost, performance, and usability.

To meet the requirements of the Central Limit Theorem, a diverse sample population (N=30) was selected, reflecting varying levels of industry experience. This is a critical consideration, as several studies have demonstrated differences in design approaches between novice and expert designers [51, 52]. Novice designers follow a simple iterative strategy of ideation, implementation, and evaluation, while experienced designers leverage their years of expertise to evaluate ideas more rigorously before implementation. This approach results in fewer iterations and less time required for refinement.

Building on the promising initial findings, a case study was conducted to quantify the benefit of this AI-driven tool. This involved incorporating 'Predictive Prototypes' into a conventional design workflow for an existing and widely used product. The Dyson Supersonic Hairdryer attachment was selected as the focus, given its availability off the shelf. Design modifications were made to the existing attachment, resulting in two distinct permutations: (1) a simulation-driven design based on pressure differential performance, and (2) a data-driven design achieved through human-AI collaboration. In the latter, the designer employed 'Predictive Prototyping' to make decisions aimed at meeting a specific design objective - improving the airflow and acoustic performance.

The introduction of GPT-RAG or 'Predictive Prototypes' demonstrates the potential to aid designers and engineers to make better-informed decisions, thereby reducing the likelihood of unnecessary iterations. While this exciting innovation offers a glimpse into a brighter future, it also presupposes that users possess a comprehensive understanding of various design aspects. Early-stage design problems, which can emerge and be resolved through design decisions, are central to this process. Essentially, with predictive prototyping, there is "one less" prototype likely to fail, and the probability of success is increased. The overarching goal should be to design for a better future - minimizing material wastage and reducing the environmental impact of iterative designs.

*Review of Research Questions*

For this review, GPTs would equate to 'Predictive Prototypes' or GPT-RAG and ChatGPT, unless specified.

1. *Do GPTs provide accurate predictions of prototype performance and cost based on input design concept-sketches, as compared to ground truth physical prototyping costs?*
   Yes, GPTs provide more accurate predictions of the actual values compared to human evaluators. The key findings are illustrated in Figures 6 and 8 and summarized in Table 2.

2. *Can GPTs accurately predict usability issues as compared to ground truth prototyping efforts?*
   Yes, GPTs demonstrate the capability to accurately predict usability issues, although human evaluators perform slightly better, likely due to their real-world experience as industry experts. In retrospect, GPTs provide a broader, more well-distributed range of predictions across key terms but exhibit less precision when compared to the ground truth derived from prototyping efforts. While humans currently excel in this area, the performance gap is not significant and may narrow further with the rapid advancements in generative AI technologies. The key findings are summarized in Figure 11.

3. *How do GPT predictions compare to individual human, and crowd-based estimations of design cost, performance, and usability?*
   GPTs tend to perform better in data-rich contexts, such as cost and performance predictions, whereas humans outperform in usability assessments, as user interactions and experiences remain highly qualitative and inherently subjective. A notable trend emerges in crowdsourcing and repeated querying, where results have shown a reduction in RMSE and an improved accuracy. This highlights an exciting opportunity to leverage multiple querying to emulate crowdsourcing, offering a valuable tool for design thinking tasks. The key findings are summarized in Table 2 and Figure 12.

4. *Does incorporation of a prototyping database significantly increase GPT prediction accuracy?*
   Yes, based on Tables 2 and 3, the p-value indicates statistical significance across the three groups. This is further supported by Figures 7 and 9, which illustrate the improved RMSE loss function when GPT is augmented with RAG.

*Limitations*

1. **Hallucinations and poor repeatability in responses**

   During the initiation phase of the experiments, several challenges arose in engineering prompts to generate meaningful insights. First, the prompts needed to be well-structured to facilitate mutual communication for both human users and GPTs, ensuring a fair comparison. Second, discrepancies in the interpretation of 2D images (i.e., labeled versus unlabeled) required manual rectification to maintain consistency across all 12 designs. Lastly, balancing the system prompt, input prompt, and data retrieval from the vector datastore proved challenging. This required multiple trial-and-error iterations to identify the most effective approach.

   Given the grey-box nature of the methodology, users had limited control over the generation process, introducing additional challenges. Common issues included overly generic responses, significant variability in repeated outputs (sometimes diverging considerably from initial predictions), and irrelevant or off-topic outputs, often referred to as "hallucinations."

   A case study involving a hairdryer attachment was relatively straightforward due to the accessibility of relevant data. However, this accessibility may not extend to more complex systems, such as satellites or rocket thrusters, where relevant data is less readily available to public. A key limitation of the research lies in the lack of reliability testing to ensure the tool's robustness for full-scale production. Furthermore, these models depend on static data, which is misaligned with the dynamic nature of the real world, presenting another significant constraint.

2. **Limitations of model and key metrics**

   During the research phase, five key metrics were identified to evaluate successful designs: cost, performance, usability, manufacturability, and production time. However, due to resource constraints, the focus was narrowed to three key metrics, resulting in significant gaps within the evaluation framework. Although initial progress has been made, the research remains an initial step towards a more comprehensive approach. Further studies could expand on the various aspects of development.

   To bridge these gaps, future efforts might prioritize quantifying manufacturability, or production costs. This will further define the suitability and best practices for using GPTs to generate production-ready designs that better align with real-world requirements, although this set of information is also not as readily available or quantifiable.

   This paper highlights the vast potential of leveraging on AI-driven tools in design but also calls attention to the iterative refinements needed to achieve industry leading solutions. Continuing to evolve this framework will unlock a transformative approach to product design.

**ACKNOWLEDGEMENTS**

This work would not have been possible without the generous support of Dyson Operations Pte Ltd, the Singapore University of Technology and Design (SUTD), and the Economic Development Board (EDB) of Singapore. I am deeply grateful to my supervisors for their invaluable guidance and expertise, which have been instrumental in bridging the gap between industrial applications and academic research, enriching both the depth and scope of this study. I also extend my sincere thanks to the rightful design project owners, including the contributors on 'Instructables.com', whose invaluable contributions to the open-source community have provided a strong foundation and inspiration to propel this research forward.

# REFERENCES
Citations and Bibliography